\documentclass{article}
\usepackage{graphicx,epsfig,rotating}

\newlength{\figwidth}
\newcommand\GeV{\ifmmode {\mathrm{\ Ge\kern -0.1em V}}\else
                   \textrm{Ge\kern -0.1em V}\fi}
\begin{document}

\setlength{\unitlength}{1mm} \thispagestyle{empty}

\begin{center}
{\Large EUROPEAN ORGANIZATION FOR NUCLEAR RESEARCH}
\end{center}

\begin{flushright}
  {\large CERN-EP/99-70 }\\
  {\large DSF  15/99} \\
  {\large May 11, 1999}\\
\end{flushright}
\vspace{2cm}

\begin{center}
{\Large \textbf{A direct evaluation of the $\Lambda_c^+$ absolute
branching ratios: a new approach }}
\end{center}

\vspace{2cm}


\begin{center}
\textbf{\large P. Migliozzi\footnote{Now also at the INFN, Sezione di Napoli, Italy}}\\[0pt]
\vspace{0.2cm} {CERN, Geneva, Switzerland}\\[0pt]
\vspace{0.5cm} \textbf{\large G. D'Ambrosio, G. Miele, P. Santorelli}\\[0pt]
\vspace{0.2cm} {Universit\`{a} Federico II and INFN, Napoli, Italy}

\begin{abstract}
  A novel method for a direct measurement of the exclusive $\Lambda_c^+$
  branching ratios is here described. The approach is based on the peculiar
  topology of the quasi-elastic charm neutrino-induced events on nucleons. The
  intrinsic potentiality of the method is thoroughly discussed using a
  \emph{perfect detector}. As an application, the statistical accuracy
  reachable with existing data sample has been estimated. From a theoretical
  point of view, such measurement provides a better understanding of the
  baryonic b-decays.
\end{abstract}
{\it Submitted to Physics Letters B}
\end{center}

\newpage
%
%

\section{Introduction}

There exists no direct measurement of $BR(\Lambda _{c}^{+}\rightarrow pK^{-}\pi
^{+})$. However, using different assumptions and experimental data, two
indirect determinations of the above branching ratio have been given
~\cite{pdg,dunietz}. These two values are quite different, and
thus it is likely that one of the procedures is incorrect.\\
Note that, since the channel $\Lambda _{c}^{+}\rightarrow pK^{-}\pi^{+}$ is
used for normalisation, a change in its branching ratio will affect most of the
$\Lambda_{c}^{+}$ exclusive decay widths.  \\ \indent Here we propose a method,
based on neutrino quasi-elastic charm production process, to directly measure
the $\Lambda _{c}^+$ branching ratios. This measurement, solving the above
puzzle, will provide new insight on the underlying hadronic physics.  \\
\indent This letter has the following structure: in Section~\ref{model} we
discuss the two different determinations of $BR(\Lambda_c^+\rightarrow
pK^-\pi^+)$ and the physics impact of a direct measurement. In
Section~\ref{chprod}, the charm production in neutrino scattering is analysed
by studying topology, kinematics and cross-sections of quasi-elastic processes
and deep inelastic reactions, which are the main source of background. The
evaluation of the accuracy achievable for these $\Lambda_{c}^{+}$-exclusive
decay widths in neutrino scattering is performed in Section~\ref{metho}. In the
last section, we give our conclusions.

%
%

\section{Model dependent extraction of $BR(\Lambda_c^+\rightarrow pK^-\pi^+)$
}
\label{model}
As stated above, two different methods to extract $BR(\Lambda
_{c}^{+}\rightarrow pK^{-}\pi ^{+})$ from the experimental data
have been proposed in literature ~\cite{pdg,dunietz}.

\noindent
\textbf{Method }$\mathcal{A}$

\noindent
The ARGUS ~\cite{argus1} and CLEO ~\cite{cleo1} experiments have measured the
quantity $BR(\bar{B}\rightarrow \Lambda _{c}^{+}X){\times} BR(\Lambda
_{c}^{+}\rightarrow pK^{-}\pi ^{+})$. Moreover, under the assumptions that
$\Gamma (\bar{B}\rightarrow \mbox{baryon}~X)$ is dominated by $\bar{B}
\rightarrow \Lambda _{c}^{+}X$, CLEO ~\cite{cleo1} and ARGUS ~\cite{argus2}
have been able to determine the branching ratio $BR(\bar{B}\rightarrow
\Lambda_{c}^{+}X)$. Thus, combining these above measurements one obtains
~\cite{pdg}
\begin{equation}
BR(\Lambda _{c}^{+}\rightarrow pK^{-}\pi ^{+})=(4.14{\pm} 0.91)\%~~.
\end{equation}
It is worth pointing out, as emphasised in ~\cite{dunietz}, that in order to
derive $BR(\bar{B} \rightarrow \Lambda _{c}^{+} X)$
\begin{itemize}
\item[i)] $\Xi_c$ and $\Omega_{c}$ production in $\bar{B}$-decays,
\item[ii)]$\bar{B}\rightarrow D^{*}\bar{N}NX$ like decay channels
\end{itemize}
have been neglected.

\noindent
\textbf{Method }$\mathcal{B}$

\noindent
A different determination of $BR(\Lambda_{c}^{+}\rightarrow pK^{-}\pi
^{+})$ is based on the independent measurements of \cite{argus4,cleo3}
\begin{equation}
\sigma (e^{+}e^{-}\rightarrow \Lambda _{c}^{+}X){\times} BR(\Lambda
_{c}^{+}\rightarrow pK^{-}\pi ^{+})\,,
\end{equation}
and \cite{argus3,cleo2}
\begin{equation}
\sigma (e^{+}e^{-}\rightarrow \Lambda _{c}^{+}X){\times} BR(\Lambda
_{c}^{+}\rightarrow \Lambda l^{+}\nu _{l})\,.
\end{equation}
Averaging over the results of
the two experiments one obtains ~\cite{pdg}
\begin{eqnarray}
R& \equiv & \frac{BR(\Lambda _{c}^{+}\rightarrow pK^{-}\pi
^{+})}{BR(\Lambda
_{c}^{+}\rightarrow \Lambda l^{+}\nu _{l})}=\frac{\sigma
(e^{+}e^{-}\rightarrow \Lambda _{c}^{+}X){\times} BR(\Lambda
_{c}^{+}\rightarrow pK^{-}\pi ^{+})}{\sigma (e^{+}e^{-}\rightarrow \Lambda
_{c}^{+}X){\times} BR(\Lambda _{c}^{+}\rightarrow \Lambda l^{+}\nu _{l})} \nonumber \\
& = & 2.40{\pm} 0.43\, .
\label{Rexpt}
\end{eqnarray}
Then the branching ratio for $\Lambda _{c}^{+}\rightarrow pK^{-}\pi ^{+}$
can be obtained from the following relation
\begin{equation}
BR(\Lambda _{c}^{+}\rightarrow pK^{-}\pi ^{+})=\frac{R{\cdot}
 f {\cdot} F}{1+\mid
V_{cd}/V_{cs}\mid ^{2}}\,
\Gamma (D^{0}\rightarrow Xl^{+}\nu _{l}) \, \tau (\Lambda _{c}^{+})\,,
\label{equa1}
\end{equation}
where
\begin{eqnarray}
 f & \equiv & \frac{BR(\Lambda _{c}^{+}\rightarrow \Lambda
l^{+}\nu _{l})}{BR(\Lambda _{c}^{+}\rightarrow X_{s}l^{+}\nu _{l})}\,,\\ F
& \equiv &  \frac{\Gamma (\Lambda _{c}^{+}\rightarrow X_{s}l^{+}\nu
_{l})}{\Gamma (D^{0}\rightarrow X_{s}l^{+}\nu _{l})}\,.
\end{eqnarray}
The ratio $f$ is obviously $<$1 since there are non-vanishing
contributions to the leading $X_{s}$ = $\Lambda $ from $\Sigma ^{{\pm} }$ $\pi
^{\mp },$ and $n\bar{K}^{0},pK^{-},$  which however are  phase space
suppressed.  The analogous fraction for the meson $D$ has been measured by
CLEO ~\cite{cleo93}
\begin{equation}
 \frac{\Gamma (D\rightarrow (K+K^{*})l^{+}\nu
_{l})}{\Gamma (D\rightarrow X_{s}l^{+}\nu _{l})}\ =\ 0.89{\pm} 0.12\, .
\end{equation}
Thus one can reasonably expect $f$ $\simeq 0.9{\pm} 0.1.$ In any case,
experiments might measure $f$ directly.\\ \indent Theoretical estimates of $F$
are based on Operator Product Expansion in the framework of the heavy quark
effective theory ~\cite{OPE}, where the amplitudes for $\Lambda
_{c}^{+}\rightarrow X_{s}l^{+}\nu _{l}$ and $D^{0}\rightarrow X_{s}l^{+}\nu
_{l}$ are predicted to have the same leading terms, while the $\mathcal{\
  O(}$1/$m_{c}^{2})$ corrections are found to be larger for $\Lambda _{c}^{+}$.
As a result $F=1.3{\pm} 0.2$ ~\cite{OPE}. Thus, if one uses for $R$ the value
of Eq.(\ref{Rexpt}) and ~\cite{pdg} for
\begin{eqnarray}
1+\mid V_{cd}/V_{cs}\mid ^{2} & = & 1.05\,, \\
\Gamma (D^0\rightarrow Xl^{+}\nu _{l}) & = & (0.163{\pm}
 0.006){\times} 10^{12}~s^{-1}\,,\\
\tau (\Lambda _{c}^{+}) & = & (0.206{\pm} 0.012){\times} 10^{-12}~s\,,
\end{eqnarray}
we obtain from Eq.( \ref{equa1})
\begin{equation}
BR(\Lambda_c^+\rightarrow pK^-\pi^+) = (7.7{\pm}1.5{\pm}2.3)\%\,.
\end{equation}
Ref. ~\cite{pdg} gives the weighted value of the two measurements $BR(\Lambda
_{c}^{+}\rightarrow pK^{-}\pi ^{+})=(5.0{\pm} 1.3)\%$, where also the
theoretical uncertainty is included. However, it is also stressed that this
number is rather arbitrary. Indeed Ref.~\cite{dunietz} advocates method
$\mathcal{B}\ $ suggesting that one should reanalyse the existing $\bar{B}$ $%
\rightarrow $ baryon sample to look for $\bar{B}\rightarrow D^{*}\bar{N}NX$
decay channels. As a result, if this interpretation is correct, then:
\begin{enumerate}
\item[i)]  heavy baryon tables would change;
\item[ii)]  measured charm counting in $b$-decay would decrease;
\item[iii)]  $\bar{B}$ $\rightarrow $ baryon decay models should be reanalysed.
\end{enumerate}

%
%
\section{Neutrino charm production}
\label{chprod}

\subsection{Neutrino quasi-elastic charm processes}
The simplest exclusive charm-production reaction is the quasi-elastic
process where a $d-$valence quark is changed into a $c-$quark, thus
 transforming the
target nucleon into a charmed baryon. Explicitly the quasi-elastic reactions
are
\begin{equation}  \label{rea1}
\nu_\mu n\rightarrow\mu^- \Lambda_c^+(2285),
\end{equation}
\begin{equation}  \label{rea2}
\nu_\mu p\rightarrow\mu^-\Sigma_c^{++}(2455),
\end{equation}
\begin{equation}  \label{rea3}
\nu_\mu n\rightarrow\mu^- \Sigma_c^{+}(2455),
\end{equation}
\begin{equation}  \label{rea4}
\nu_\mu p\rightarrow\mu^- \Sigma_c^{*++}(2520),
\end{equation}
\begin{equation}  \label{rea5}
\nu_\mu n\rightarrow\mu^- \Sigma_c^{*+}(2520).
\end{equation}
In literature there are
two classes of models which try to describe the
processes (\ref{rea1})-(\ref{rea5}).
The first class \cite{mode1}-\cite{mode6} is based
on the $SU(4)$ flavour symmetry, but since $SU(4)$ is badly broken,
the parameters (axial and vectorial mass)
entering the cross-section formula cannot be reliably predicted.
\\ \indent
A different approach \cite{kova} is based on the
Bloom-Gilman~\cite{bloom} local duality in $\nu N$ scattering modified on the
basis of QCD~\cite {deru1,deru2}.
\\ \indent
The cross-sections predicted by the different models, assuming a neutrino
energy of $10~GeV$, are shown in Table~\ref{tab:crosspred}.
As we can see, these predictions can even differ by one order of magnitude.
From Fig. 3 in Refs. \cite{mode2} and \cite{kova} we note that
the total cross-section in both
models is almost flat for a neutrino energy larger than $8 \ GeV$.

\begin{table}[tbp]
\begin{center}
{\small
\begin{tabular}{||c|c|c|c|c|c||}
\hline
$\sigma(10^{-39}~cm^2)\backslash$ Model & F.R.~\cite{mode1} & S.L.~\cite{mode2}
& A.K.K.~\cite{mode3,mode4,mode5} & A.G.Y.O.~\cite{mode6} & K.~\cite{kova} \\
\hline\hline
$\nu_\mu p\rightarrow\mu^-\Sigma_c^{++}$ & 0.02 & 0.9 & 0.8 & 0.1 & 0.3 \\
\hline
$\nu_\mu p\rightarrow\mu^-\Sigma_c^{*++}$ & 0.06 & 1.6 & 1.0 & 0.06 & - \\
\hline
$\nu_\mu n\rightarrow\mu^-\Lambda_c^{+}$ & 0.1 & 2.3 & 4.1 & 0.3 & 0.5 \\
\hline
$\nu_\mu n\rightarrow\mu^-\Sigma_c^{+}$ & 0.01 & 0.5 & - & 0.06 & 0.15 \\
\hline
$\nu_\mu n\rightarrow\mu^-\Sigma_c^{*+}$ & 0.03 & 0.8 & - & 0.03 & - \\
\hline\hline
\end{tabular}
}
\end{center}
\caption{Predicted quasi-elastic charm production cross-section assuming a
neutrino energy of $10~GeV$.}
\label{tab:crosspred}
\end{table}
\indent
Only one measurement of the neutrino quasi-elastic charm production
cross-section exists. The E531 experiment~\cite{e531}, based on a sample of 3
events and using a neutrino beam with an average energy of $\sim 22~GeV$,
measured the $\Lambda_c^+$ quasi-elastic production cross-section to be:
$$\sigma(\nu_\mu
n\rightarrow\mu^-\Lambda_c^+)=0.37_{-0.23}^{+0.37}{\times}10^{-39}~cm^2$$
\vspace{-0.2cm}
$$\mathcal{R}\equiv\frac{\sigma(\nu_\mu
  n\rightarrow\mu^-\Lambda_c^+)}{\sigma(\nu_\mu
  N\rightarrow\mu^-X)}=0.3^{+0.3}_{-0.2}\%$$

Despite the large statistical error,
this measurement is clearly inconsistent with the predictions of Refs.
~\cite{mode2}-\cite{mode5}, while the agreement is fair for
~\cite{mode1,mode6,kova}. An average value of the cross-sections predicted by
Refs.~\cite {mode1,mode6,kova} has been used to have a rough estimate of the
expected number of events, see Table~\ref{tab:crossused}.
\\
\begin{table}[tbp]
\begin{center}
{\small
\begin{tabular}{||c|c|c|c||}
\hline
Reaction & $\sigma(10^{-39}~cm^2)$ & $\mathcal{R}(\%)$ & Expected events \\
\hline\hline
$\nu_\mu p\rightarrow\mu^-\Sigma_c^{++}$ & 0.14 & $0.14$ & 1400 \\ \hline
$\nu_\mu p\rightarrow\mu^-\Sigma_c^{*++}$ & 0.06 & $0.06$ & 600 \\ \hline
$\nu_\mu n\rightarrow\mu^-\Lambda_c^{+}$ & 0.3 & $0.3$ & 3000 \\ \hline
$\nu_\mu n\rightarrow\mu^-\Sigma_c^{+}$ & 0.07 & $0.07$ & 700 \\ \hline
$\nu_\mu n\rightarrow\mu^-\Sigma_c^{*+}$ & 0.03 & $0.03$ & 300 \\ \hline
All & 0.6 & $0.6$ & 6000 \\ \hline\hline
\end{tabular}
}
\end{center}
\caption{Quasi-elastic charm production cross-section and its contribution
to the total charged-current neutrino cross-section. In the last column the
expected number of events, assuming a starting sample of 1 million
charge-current neutrino-induced events, is shown.}
\label{tab:crossused}
\end{table}

%
%
\label{topo}

\subsection{Topology of the quasi-elastic events}
\begin{figure}[htbp]
\begin{center}
\rotatebox{0}{\ \resizebox{0.7\textwidth}{!}{\ \includegraphics{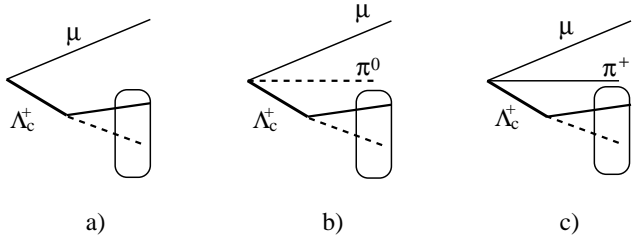} }}
\end{center}
\caption{Topology of the quasi-elastic charm neutrino induced events in the
case of the reaction a) $\nu_\mu n\rightarrow\mu^-\Lambda_c^+$, b) $\nu_\mu
n\rightarrow\mu^-\Sigma_c^+(\Sigma_c^{*+})$ and c) $\nu_\mu
p\rightarrow\mu^-\Sigma_c^{++}(\Sigma_c^{*++})$. The particles inside the
ellipse represent the $\Lambda_c^+$ decay products.}
\label{fi:topo}
\end{figure}

In the quasi-elastic processes (\ref{rea1})-(\ref{rea5}), besides the charmed
baryon, only a muon is produced at the interaction point (primary vertex). For
reaction~(\ref{rea1}) we expect the topology shown in Fig.~\ref{fi:topo}a),
namely, a muon track plus the $\Lambda_c^+$ immediately decaying. For
reactions~(\ref{rea2}) and (\ref{rea4}), since the produced $ \Sigma_c^{++}$
strongly decays into a $\Lambda_c^+$ and a $ \pi $, we expect three charged
particles at the primary vertex as shown in Fig.~\ref{fi:topo}c).  For
reactions~(\ref{rea3}) and (\ref{rea5}) we expect a number of charged particles
produced at the primary vertex equal to the one of reaction~(\ref{rea1}), plus
a neutral pion produced in the $\Sigma_c^+$ decay (see Fig.~\ref{fi:topo}b)).

Indeed, these events are characterised by a peculiar topology: two charged
tracks produced at the primary vertex, with $\Sigma_c$, if present, decaying
strongly. In any case, the final charmed baryon can only be a $\Lambda_c^+$.
This feature is very important and will be exploited heavily in the following.

%
%
\label{kine}

\subsection{Kinematics of the quasi-elastic reaction}

\begin{figure}[htbp]
\begin{center}
\rotatebox{0}{\ \resizebox{0.5\textwidth}{!}{\ \includegraphics{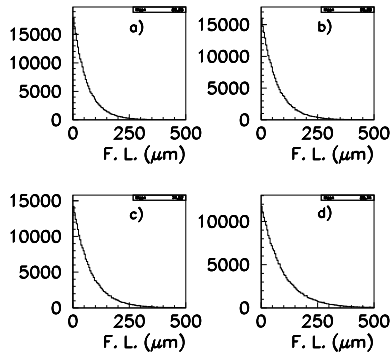} }%
}
\end{center}
\caption{Flight length distribution for different $q^2$ values: a)~$%
q^2=0.1~GeV^2$, b)~$q^2=0.5~GeV^2$, c)~$q^2=1~GeV^2$, d)~$q^2=2~GeV^2$.}
\label{fi:declen}
\end{figure}

Let us assume, for simplicity, the initial nucleon at rest. In this frame,
given the transferred 4-momentum squared $q^2$, the energy of the charmed
baryon is
\begin{equation}  \label{enebar}
E_C = \frac{q^2+m_C^2+m_N^2}{2m_N},
\end{equation}
where $m_C$ is its mass and $m_N$ is the mass of the struck nucleon.  The
$q^2-$di\-stri\-bu\-tion, at a fixed neutrino energy, $(1/\sigma)d\sigma/dq^2$
is in general model dependent ~\cite{mode1}-\cite{kova}. Nevertheless, despite
the large disagreement in the theoretical total cross-sections, the
predicted $q^2-$distributions are very similar and lie mostly in the range $
0<q^2<2~GeV^2$.  We study the kinematics of the quasi-elastic process at
different values
of $q^2$ integrating over the neutrino energy spectrum\footnote{%
  As an example of neutrino beam we considered the CERN-SPS WANF neutrino beam
  ~\cite{wanf2}, which has an average energy of about of $27~GeV$.}.  As
reference values of $q^2$ we use (0.1, 0.5, 1, 2)~$GeV^2$.  \\ \indent The
flight length and the decay kinematics of the $\Lambda_c^+$ have been evaluated
using the package PYTHIA5.7/JETSET7.4~\cite{jetset}. The flight length
distributions for four different $q^2$ values are shown in
Fig.~\ref{fi:declen}. From this figure we can see that, because of the small
$\gamma$ factor, almost all the $\Lambda_c^+$ decay within $500~\mu m$ from the
production point. This parameter is crucial when we start thinking of a
detector able to \emph{see} the topologies shown in Fig.~\ref{fi:topo}.

%
%
\label{deeppro}

\subsection{Deep inelastic charm processes}

Charmed hadrons can be produced in deep inelastic neutrino interactions
through the reaction $\nu_\mu N \rightarrow \mu^- C X$, where $C= D^0,
D^+, D_s^+, \Lambda_c^+$ and $X$ is purely hadronic. Using the CERN-SPS WANF
neutrino beam~\cite{wanf2}, $(57{\pm}5)\%$ of the charmed particles
produced in deep inelastic interactions is neutral. The rest can be
split as follows: $(19{\pm}4)\%~D^+$, $(12{\pm}4)\%~D_s^+$ and
$(12{\pm}3)\%~\Lambda_c^+$.  For details about the charm production
fractions in neutrino interactions we refer to~\cite{bolton}.
\\ \indent
Through a Monte Carlo simulation, we also checked that $(15.48{\pm}0.09)\%$
of the deep inelastic events with a charmed hadron in the final state has a
topology similar to those shown in Fig.~\ref{fi:topo}a) and Fig.~\ref
{fi:topo}b) and $(8.43{\pm}0.04)\%$ to the one in Fig.~\ref{fi:topo}c).

\subsection{Quasi-elastic versus deep inelastic charm events}

\begin{figure}[htbp]
\begin{center}
\rotatebox{0}{\ \resizebox{0.5\textwidth}{!}{\ \includegraphics{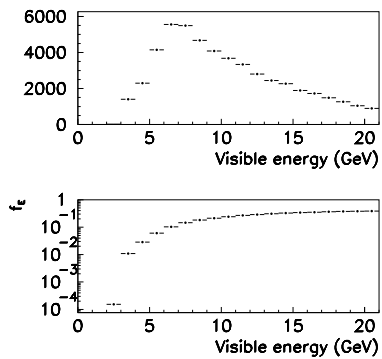} }%
}
\end{center}
\caption{{Top: Visible energy distribution of deep inelastic charm events
    with a vertex topology similar to the one shown in Fig.~\ref{fi:topo}a) and
    Fig.~\ref{fi:topo}b). Bottom: fraction of deep inelastic charm events
    faking a quasi-elastic topology as a function of the energy cut.}}
\label{fi:enevis}
\end{figure}

A suitable variable to discriminate quasi-elastic from deep inelastic events is
the visible hadronic energy, see Fig.~\ref{fi:enevis}.  According to
Eq.~(\ref{enebar}), a large fraction of the visible energy in quasi-elastic
events is expected to lie in the range $(3\div5)~GeV$. It is, indeed, possible
to achieve a high rejection power against deep inelastic events keeping high
efficiency for quasi-elastic ones. For example, the cut $E_{vis}\leq5~GeV$
rejects $(97.1{\pm}0.1)\%$ of the deep inelastic processes keeping more than
$90\%$ of quasi-elastic events.  \\ \indent It is important to have high
rejection power against deep inelastic charm production due to its large
cross-section~\cite{e531} ($\sigma(\nu_\mu N
\rightarrow \mu^- C X)/\sigma(\nu_\mu N \rightarrow \mu^-X)\sim 5\%$%
\footnote{%
  This result includes also the production of neutral charmed particles.},
$\mathcal{R}=0.6\%$) and to the fact that not only $\Lambda_c^+$, but also
$D^+$ and $D^+_s$ are produced. In fact, $D^+$ or $D^+_s$ hadrons, wrongly
identified, could bias the determination of the $\Lambda_c^+$ branching ratios.
The quasi-elastic sample contamination, which comes from deep inelastic events,
can be written using Table~\ref{tab:crosspred} as
\begin{equation}
\varepsilon_{fake} = \frac{\sigma(\nu_\mu N\rightarrow\mu^- C X)}{
\sigma(\nu_\mu N\rightarrow\mu^-X)}{\times}\frac{1}{\mathcal{R}}{\times}
f_{fake}{\times} f_{E<5GeV}{\times} f_{(D^+orD^+_s)}\,.
\end{equation}
The factor $f_{fake}(=(23.9{\pm}0.1)\%)$ denotes the fraction of deep
inelastic events which fakes a quasi-elastic topology,
$f_{E<5GeV}(=(2.9{\pm}0.1)\%)$ is the fraction of the deep inelastic
events which survives to the energy cut, and $f_{(D^+or
D^+_s)}(=(8.2{\pm}0.4)\%)$ is the fraction, among \textit{fake}
quasi-elastics, with a charmed meson in the final state. Hence, the
contamination is $\varepsilon_{fake}\simeq 0.5\%$. The relative error on
$\varepsilon_{fake}$ as a function of the relative error on
$\mathcal{R}$ is shown in Table~\ref{tab:erro}.

\begin{table}[tbp]
\begin{center}
{\small
\begin{tabular}{||c|c|c||}
\hline
$\Delta\mathcal{R}/\mathcal{R}$ & $\sigma_{fake}/\varepsilon_{fake}$ & $%
\varepsilon_{fake}$ \\ \hline\hline
$10\%$ & $19\%$ & $(0.45{\pm}0.09)\%$ \\ \hline
$30\%$ & $34\%$ & $(0.45{\pm}0.15)\%$ \\ \hline
$50\%$ & $53\%$ & $(0.45{\pm}0.24)\%$ \\ \hline
$100\%$ & $101\%$ & $(0.45{\pm}0.45)\%$ \\ \hline
$200\%$ & $201\%$ & $(0.45{\pm}0.90)\%$ \\ \hline
$500\%$ & $500\%$ & $(0.45{\pm}2.25)\%$ \\ \hline\hline
\end{tabular}
}
\end{center}
\caption{The relative and absolute errors on $\varepsilon_{fake}$ are shown as
  a function of the relative error on $\mathcal{R}$.}
\label{tab:erro}
\end{table}

%
%

\section{Description of the method}
\label{metho}
Due to the peculiar topology and kinematics of the quasi-elastic events an
almost pure sample of $\Lambda_c^+$, with a small contamination of $D^+$ and
$D^+_s$ produced in deep inelastic events, can be selected. No model dependent
information is used to determine the number of $\Lambda_c^+$. The
contamination of $D^+$ and $D^+_s$ from deep inelastic events can be taken into
account assigning it as relative systematic error on the branching ratios. We
assumed the relative systematic error to be
$\varepsilon_{fake}+3\sigma_{fake}$. The normalisation to determine the
$\Lambda_c^+$ absolute branching ratios is simply given by the number of events
with a vertex topology consistent with Fig.~\ref{fi:topo}. We want to stress,
here, that no model dependent information is used to determine the
normalisation. The little knowledge we have about the quasi-elastic charm
production cross-section, which is model dependent unless measured,
plays a role
only in the evaluation of the deep inelastic contamination, namely the
systematic error. It is worth noticing that, even if the ratio $\mathcal{R}$
is known with an uncertainty of $500\%$\footnote{Nevertheless, this is not the
  case. We recall that E531~\cite{e531}, with only 3 events observed, measured
  $\mathcal{R}$ with an accuracy of $100\%$.}, the relative systematic error on
the branching ratios is $\sim7.2\%$ (see Table~\ref{tab:erro}).

%
%

\subsection{Measurement accuracy in a \emph{perfect detector}}

\label{sensi}
The \emph{perfect detector} by definition has the following features:

\begin{itemize}
\item  vertex topology measurement with $100\%$ efficiency
resolution for charmed baryon decays;

\item  infinite energy resolution;

\item  particle identification capability with infinite resolution.
\end{itemize}

\begin{table}[tbp]
\begin{center}
{\small
    \begin{tabular}{||c|c|c|c|c||}
      \hline
      Channel & PDG BR~\cite{pdg} & $\Delta BR$ &$\Delta BR$  & $\Delta BR$ \\
       &  & ($\frac{\Delta\mathcal{R}}{\mathcal{R}}=10\%$) & ($\frac{\Delta\mathcal{R}}{\mathcal{R}}=100\%$)& ($\frac{\Delta\mathcal{R}}{\mathcal{R}}=500\%$)\\
      \hline
      \hline
      $\Lambda_c^+\rightarrow p\bar{K}^0$ & $(2.5{\pm}0.7)\%$ & $({\pm}0.2{\pm}0.02)\%$& $({\pm}0.2{\pm}0.05)\%$& $({\pm}0.2{\pm}0.2)\%$\\
      \hline
      $\Lambda_c^+\rightarrow p K^-\pi^+$ & $(5.0{\pm}1.3)\%$ & $({\pm}0.3{\pm}0.04)\%$& $({\pm}0.3{\pm}0.09)\%$& $({\pm}0.3{\pm}0.4)\%$\\
      \hline
      $\Lambda_c^+\rightarrow p\bar{K}^0\eta$ & $(1.3{\pm}0.4)\%$ & $({\pm}0.1{\pm}0.01)\%$ & $({\pm}0.1{\pm}0.02)\%$& $({\pm}0.1{\pm}0.09)\%$\\
      \hline
      $\Lambda_c^+\rightarrow p\bar{K}^0\pi^+\pi^-$ & $(2.4{\pm}1.1)\%$ &
      $({\pm}0.2{\pm}0.02)\%$& $({\pm}0.2{\pm}0.04)\%$& $({\pm}0.2{\pm}0.2)\%$\\
      \hline
      $\Lambda_c^+\rightarrow p K^-\pi^+\pi^0$ & $(4.7{\pm}1.3)\%$ & $({\pm}0.3{\pm}0.04)\%$& $({\pm}0.3{\pm}0.08)\%$& $({\pm}0.3{\pm}0.3)\%$\\
      \hline
      $\Lambda_c^+\rightarrow \Lambda \pi^+\pi^0$ & $(3.6{\pm}1.3)\%$ & $({\pm}0.2{\pm}0.03)\%$& $({\pm}0.2{\pm}0.06)\%$& $({\pm}0.2{\pm}0.3)\%$\\
      \hline
      $\Lambda_c^+\rightarrow \Lambda \pi^+\pi^+\pi^-$ & $(3.3{\pm}1.0)\%$ & $({\pm}0.2{\pm}0.02)\%$& $({\pm}0.2{\pm}0.06)\%$& $({\pm}0.2{\pm}0.2)\%$\\
      \hline
      $\Lambda_c^+\rightarrow \Lambda \pi^+\eta$ & $(1.7{\pm}0.6)\%$ & $({\pm}0.2{\pm}0.01)\%$& $({\pm}0.2{\pm}0.03)\%$& $({\pm}0.2{\pm}0.1)\%$\\
      \hline
      $\Lambda_c^+\rightarrow \Sigma^+ \pi^0$ & $(1.0{\pm}0.3)\%$ & $({\pm}0.1{\pm}0.01)\%$& $({\pm}0.1{\pm}0.02)\%$& $({\pm}0.1{\pm}0.07)\%$\\
      \hline
      $\Lambda_c^+\rightarrow \Sigma^+ \pi^+\pi^-$ & $(3.4{\pm}1.0)\%$ & $({\pm}0.2{\pm}0.02)\%$& $({\pm}0.2{\pm}0.06)\%$& $({\pm}0.2{\pm}0.2)\%$\\
      \hline
      $\Lambda_c^+\rightarrow \Sigma^- \pi^+\pi^+$ & $(1.8{\pm}0.8)\%$ & $({\pm}0.2{\pm}0.01)\%$ & $({\pm}0.2{\pm}0.03)\%$& $({\pm}0.2{\pm}0.1)\%$\\
      \hline
      $\Lambda_c^+\rightarrow \Sigma^0 \pi^+\pi^0$ & $(1.8{\pm}0.8)\%$ & $({\pm}0.2{\pm}0.01)\%$& $({\pm}0.2{\pm}0.03)\%$& $({\pm}0.2{\pm}0.1)\%$\\
      \hline
      $\Lambda_c^+\rightarrow \Sigma^0 \pi^+\pi^+\pi^-$ & $(1.1{\pm}0.4)\%$ & $({\pm}0.1{\pm}0.01)\%$& $({\pm}0.1{\pm}0.02)\%$& $({\pm}0.1{\pm}0.08)\%$\\
      \hline
      $\Lambda_c^+\rightarrow \Sigma^+ \pi^+\pi^-\pi^0$ & $(2.7{\pm}1.0)\%$ & $({\pm}0.2{\pm}0.02)\%$& $({\pm}0.2{\pm}0.05)\%$& $({\pm}0.2{\pm}0.2)\%$\\
      \hline
      $\Lambda_c^+\rightarrow \Lambda \mu^+\nu_\mu$ & $(2.0{\pm}0.7)\%$ & $({\pm}0.2{\pm}0.01)\%$& $({\pm}0.2{\pm}0.04)\%$& $({\pm}0.2{\pm}0.1)\%$\\
      \hline
      $\Lambda_c^+\rightarrow \Lambda e^+\nu_e$ & $(2.1{\pm}0.7)\%$ & $({\pm}0.2{\pm}0.01)\%$&  $({\pm}0.2{\pm}0.04)\%$& $({\pm}0.2{\pm}0.2)\%$\\
      \hline
      \hline
    \end {tabular}
}
\end{center}
\caption{Statistical and systematic accuracy achievable in the
  determination of the $\Lambda_c^+$ absolute branching ratios, assuming a
  collected statistics of $10^6~\nu_\mu$ charged-current events, as a function
  of the relative error on $\mathcal{R}$. The central values are taken from
ref. \cite{pdg}.}
\label{tab:brscena1}
\end{table}

In Table~\ref{tab:brscena1} the expected accuracy on the determination
of the $\Lambda_c^+$ branching ratios as a function of the relative
error on $\mathcal{R}$ is shown. To compute the expected number of
events in each decay channel we used the central values (shown in
Table~\ref{tab:brscena1} together with their errors) given by the
Particle Data Group~\cite{pdg}. From this Table we can see that, even
assuming very large (unrealistic) systematic error, it is still possible
to discriminate among methods $\mathcal{A}$ and $\mathcal{B}$ discussed in
Section~\ref{model}. In Table~\ref{tab:brscena2} the achievable accuracy
on the absolute $BR$ determination as a function of the collected
charged-current statistics is shown.

\begin{table}[tbp]
\begin{center}
{\small
\begin{tabular}{||c|c|c|c||}
\hline
Channel & $N_\mu=10^6$ & $N_\mu=10^5$ & $N_\mu=10^4$ \\
\hline\hline
$\Lambda_c^+\rightarrow p\bar{K}^0$ & $({\pm}0.2{\pm}0.05)%
\% $ & $({\pm}0.6{\pm}0.05)\%$ & $({\pm}1.8{\pm}0.05)\%$ \\ \hline
$\Lambda_c^+\rightarrow p K^-\pi^+$ & $({\pm}0.3{\pm}0.09)%
\% $ & $({\pm}0.9{\pm}0.09)\%$ & $({\pm}2.8{\pm}0.09)\%$ \\ \hline
$\Lambda_c^+\rightarrow p\bar{K}^0\eta$ & $%
({\pm}0.1{\pm}0.02)\%$ & $({\pm}0.5{\pm}0.02)\%$ & $({\pm}1.7{\pm}0.02)\%$ \\
\hline
$\Lambda_c^+\rightarrow p\bar{K}^0\pi^+\pi^-$ & $%
({\pm}0.2{\pm}0.04)\%$ & $({\pm}0.6{\pm}0.04)\%$ & $({\pm}1.0{\pm}0.04)\%$ \\
\hline
$\Lambda_c^+\rightarrow p K^-\pi^+\pi^0$ & $%
({\pm}0.3{\pm}0.08)\%$ & $({\pm}0.9{\pm}0.08)\%$ & $({\pm}2.8{\pm}0.08)\%$ \\
\hline
$\Lambda_c^+\rightarrow \Lambda \pi^+\pi^0$ & $%
({\pm}0.2{\pm}0.06)\%$ & $({\pm}0.8{\pm}0.06)\%$ & $({\pm}2.3{\pm}0.06)\%$ \\
\hline
$\Lambda_c^+\rightarrow \Lambda \pi^+\pi^+\pi^-$ & $%
({\pm}0.2{\pm}0.06)\%$ & $({\pm}0.7{\pm}0.06)\%$ & $({\pm}2.3{\pm}0.06)\%$ \\
\hline
$\Lambda_c^+\rightarrow \Lambda \pi^+\eta$ & $%
({\pm}0.2{\pm}0.03)\%$ & $({\pm}0.5{\pm}0.03)\%$ & $({\pm}1.7{\pm}0.03)\%$ \\
\hline
$\Lambda_c^+\rightarrow \Sigma^+ \pi^0$ & $%
({\pm}0.1{\pm}0.02)\%$ & $({\pm}0.4{\pm}0.02)\%$ & $({\pm}1.0{\pm}0.02)\%$ \\
\hline
$\Lambda_c^+\rightarrow \Sigma^+ \pi^+\pi^-$ & $%
({\pm}0.2{\pm}0.06)\%$ & $({\pm}0.7{\pm}0.06)\%$ & $({\pm}2.3{\pm}0.06)\%$ \\
\hline
$\Lambda_c^+\rightarrow \Sigma^- \pi^+\pi^+$ & $%
({\pm}0.2{\pm}0.03)\%$ & $({\pm}0.5{\pm}0.03)\%$ & $({\pm}1.7{\pm}0.03)\%$ \\
\hline
$\Lambda_c^+\rightarrow \Sigma^0 \pi^+\pi^0$ & $%
({\pm}0.2{\pm}0.03)\%$ & $({\pm}0.5{\pm}0.03)\%$ & $({\pm}1.7{\pm}0.03)\%$ \\
\hline
$\Lambda_c^+\rightarrow \Sigma^0 \pi^+\pi^+\pi^-$ & $%
({\pm}0.1{\pm}0.02)\%$ & $({\pm}0.4{\pm}0.02)\%$ & $({\pm}1.0{\pm}0.02)\%$ \\
\hline
$\Lambda_c^+\rightarrow \Sigma^+ \pi^+\pi^-\pi^0$ & $%
({\pm}0.2{\pm}0.05)\%$ & $({\pm}0.7{\pm}0.05)\%$ & $({\pm}2.2{\pm}0.05)\%$ \\
\hline
$\Lambda_c^+\rightarrow \Lambda \mu^+\nu_\mu$ & $%
({\pm}0.2{\pm}0.04)\%$ & $({\pm}0.6{\pm}0.04)\%$ & $({\pm}1.8{\pm}0.04)\%$ \\
\hline
$\Lambda_c^+\rightarrow \Lambda e^+\nu_e$ & $%
({\pm}0.2{\pm}0.04)\%$ & $({\pm}0.6{\pm}0.04)\%$ & $({\pm}1.8{\pm}0.04)\%$ \\
\hline\hline
\end{tabular}
}
\end{center}
\caption{Accuracy achievable ($\Delta BR$), assuming a 100\% error on  $\mathcal{R}$,
as a function of the collected charged-current statistics.}
\label{tab:brscena2}
\end{table}

\subsection{Measurement accuracy with statistics of present and future experiments}

Among the neutrino experiments which are currently taking data or
analysing data, CHORUS~\cite{chorus}, which uses nuclear emulsions as
a target, has an adequate spatial resolution to fully exploit the topologies
shown in Fig.~\ref{fi:topo}. Starting from a sample of about of 500000
charged-current events, it is estimated that $\sim 350000$ events will be
analysed in the emulsions~\cite{morio}. Assuming a 50\% efficiency to detect
the $\Lambda_c^+$ decay and taking into account that $\mathcal{R}=0.6\%$,
a statistics of $\sim 1000$ quasi-elastic events can be expected. Due to the
good muon identification of the electronic detector and the emulsion capability
in identifying electrons,  CHORUS is well suited to study $\Lambda_c^+$
semi-leptonic decays. Assuming 100\% error on $\mathcal{R}$ and the PDG central
value \cite{pdg}, the following accuracy could be achieved:
\begin{equation}
\Delta BR(\Lambda_c^+\rightarrow \Lambda \mu^+\nu_\mu) = (\pm0.44\mid_{stat}\pm0.01\mid_{sys})\%
\end{equation}

\begin{equation}
\Delta BR(\Lambda_c^+\rightarrow \Lambda e^+\nu_e) = (\pm0.45\mid_{stat}\pm0.01\mid_{sys})\%
\end{equation}

A measurement with higher sensitivity could be performed exposing a dedicated
detector, whose feasibility study has not yet been worked out, at the future
neutrino beams from muon storage rings~\cite{muon}. Such beams could provide
$\mathcal{O}(10^6)~\nu_\mu - CC$ $events/year$ in a $10~kg$ fiducial mass
detector, $1~km$ away from the neutrino source, allowing for a strong reduction
of the statistical uncertainty on the $\Lambda_c$ branching ratios, see Table~\ref{tab:brscena2}.
%
%

\section*{Conclusions}
We have presented a new method for a direct evaluation of the $\Lambda_c^+$
branching ratios. We have also stressed that this has very interesting
consequences for our understanding of baryon production in charm decays. Even
just a good determination of the $\Lambda_{c}^{+}$ semi-leptonic exclusive
decay widths would be sufficient: indeed from Eq.(\ref{Rexpt}) we would
determine $BR(\Lambda_{c}^{+}\rightarrow pK^{-}\pi ^{+})$, very important phenomenologically
as pointed out particularly in Section~\ref{model}. We have also shown that,
already with existing data, it should be possible to perform a direct measurement of $\Lambda_{c}^{+}$
exclusive decay widths, which is lacking to date.
%


%
%


\end{document}